\documentclass{aa}
\usepackage{graphicx}
\usepackage{txfonts}
\usepackage{natbib}

\begin{document}

\title{Photospheric Hanle diagnostic of weak magnetic dipoles in stars.}
\author{{\sc A.~L{\'o}pez Ariste}\inst{1}, A. Asensio Ramos \inst{2}, C. Gonz\'{a}lez Fern\'{a}ndez\inst{3}}
\institute{THEMIS - CNRS UPS 853. C/ V\'{\i}a L\'actea s/n. 38200. La Laguna. Spain.  \and
Instituto de
  Astrof\'{\i}sica de Canarias. C/ V\'{\i}a L\'actea s/n. 38200. La
  Laguna. Tenerife. Spain. \and
Departamento de F\'{\i}sica, Ingenier\'{\i}a de Sistemas y Teor\'{\i}a de la Se\~{n}al, Universidad de Alicante, 
Apdo. 99, E03080 Alicante, Spain.}
\offprints{Arturo.Lopez@themis.iac.es}
\date{Received ; accepted}
\begin{abstract} 
{Diagnostic techniques for stellar magnetic fields based upon spectropolarimetry}{We propose and explore a new technique based upon the linear
polarization emitted in Hanle-sensitive lines in disk-integrated stars where a dipolar magnetic field breaks the rotational symmetry of the resonance scattering polarization}
{A star with a simple dipolar field and a 1-0 spectral line are used to compute polarization amplitudes and angles.}
{Predicted amplitudes are low but within reach of present instruments}{A new application of the Hanle effect is proposed and analyzed, a tool 
that allows measuring  of some of the weakest stellar
magnetic fields.}
\end{abstract}
\authorrunning{L\'opez Ariste et al.}
\titlerunning{Hanle diagnostic of stellar dipoles}
\maketitle

\section{Introduction}
Magnetic fields accompany stars from the cradle to the grave: they have been detected in hot molecular
 cores \citep{girart_magnetic_2009}, they are known to be 
responsible for the electromagnetic emission of pulsars \citep{ferrario_origin_2008}
 and they also have long 
been detected in other stellar remnants such as white dwarfs \citep{kemp_discovery_1970}. Magnetic fields influence 
diverse
physical processes, such as mass accretion and loss, stellar rotation or elemental diffusion, that
critically affect the evolution of their host stars. And yet we lack a proper explanation of their origin, and 
our understanding of their evolution along the HR diagram is still fragmentary.

Outside our Sun, the vast majority of the direct information (i.e. excluding indirect indexes of magnetic 
activity such as X-ray emission) we possess on stellar magnetic fields comes from the exploitation of the various 
observational features associated with the Zeeman effect. These appear in intensity (line broadening) 
and in polarized light (as signatures in linear and circular polarization). 
Magnetic line broadening is very weak even 
for strong fields, and competes with other processes such as thermal and rotational broadening. Polarization 
measurements, on the other hand are much richer in information and sensitive to weaker fields (of a few G, according to 
\citealp{donati_magnetic_2009}) although they often require the use of 
specialized instrumentation such as Balmer-line Zeeman analysers \citep{angel_magnetic_1970}
 or high resolution echelle 
spectropolarimeters (as those developed by \citealp{semel_zeeman-doppler_1993} or \citealp{donati_espadons:_2003}). But these polarimetric 
instruments also open the door to diagnostics based upon other physical phenomena also affecting polarization and dependent on 
magnetic fields.
 With this paper we aim to provide the theoretical background to a new tool for the detection of even weaker magnetic 
fields, based on the measurement of the Hanle effect, that will complement in sensitivity and information 
content the existent techniques.

\section{Disk-integrated polarization from a magnetic dipole}
Resonance scattering polarization is the result of an anisotropic illumination of emitting atoms. Spectral lines formed in a
 stellar atmosphere can show resonance scattering polarization if their region of
 formation is sufficiently high for the radiation field to be anisotropic. Limb-darkening contributes to increase the anisotropy. The 
emitted light is linearly polarized, and its degree of polarization depends on the scattering angle. In the classical approximation, the
 polarization will be zero in forward scattering and maximum at 90 degrees scattering. Resonance scattering is however a quantum 
phenomenon and the classical view provides only a rough approximation to the actual degree of polarization, though one that serves 
the purpose of
 illustrating the main issue with the stellar observation of resonance scattering polarization. The linear polarization emitted will 
be maximum at the limb of the stellar disc and zero at the center. From pure symmetry considerations, the direction of polarization will
 be tangent to the stellar limb at every position angle. As we integrate over all  position angles, the resultant polarization will be 
zero if the star presents a spherical aspect.

Rotationally deformated stars, non-spherical stellar envelopes or winds constitute examples in which the spherical symmetry has been 
broken and the integral over all the emitting points may end up being different from zero. These cases have been extensively studied in 
the past by Ignace and collaborators \citep{ignace_hanle_2001,ignace_scattering_2008, ignace_scattering_2009, ignace_circumstellar_2008}. 

But here we want to concentrate on the case of stellar photospheres with spherical symmetry. And with that aim we turn our attention to the
fact that 
resonance scattering polarization may also be modified by magnetic fields. This is the so-called Hanle effect. It is commonly used in 
solar physics to measure magnetic fields in prominences \citep{casini_magnetic_2003} and in the quiet solar 
photosphere \citep{stenflo_hanle_1982,faurobert_investigation_2001}.
Generally speaking, the Hanle effect induces 
two primary modifications into  resonance scattering polarization arising from a single scattering event: it depolarizes the emitted
light and rotates its plane of polarization. 
The amount of depolarization and rotation will depend on the strength of the field, but also on the geometry of the three distinctive 
directions present in the problem: the direction of preferential illumination of the atom (usually the local vertical), the magnetic 
field and the line of sight.
Calculating the polarization in a spectral line from a nonresolved star requires then the addition of all the  scattering events in its atmosphere as the
line-of-sight changes in direction respect to the two other directions.  
 The theory describing those dependencies has been completely developed for spectrally resolved lines \citep{landi_deglinnocenti_polarization_2004}.
We shall in this work use the full theory, but in the following paragraphs and in the sake of illustration, we will voluntarily simplify
our description of the  Hanle effect and limit it
to just those two modifications of the polarization generated at every single scattering event.
Even further, rather than going into a detailed description of those two modifications we shall first take the opposite direction and make use of a 
widely known  
geometric property of the Hanle effect: a magnetic field perfectly aligned with the direction of illumination produces no effect on the 
resonance scattering polarization. Such simplifications capture still the essence of the Hanle effect in the case under study without burdening it 
with unnecesary details, unneeded at this point and incorporated nevertheless in the computations below.

It is time now to set the example that will explain our proposed diagnostic. Let us suppose a spherical star rotating around an axis 
in the plane of the sky, that is with the equator seen edge-on. The density and temperature of the star are spherically symmetric and it
presents limb darkening. We shall be interested in relatively weak photospheric lines for which the last scattering approximation applies \cite{stenflo_hanle_1982}
Let us suppose a global dipolar field with strength $B_0$ at the stellar surface. We will further suppose in our first example that 
the dipole is perfectly aligned with the rotation axis.  For rays from most of the equator line the dipole field will primarily induce a depolarization 
of the background
 resonance scattering polarization. 
At the poles however the dipolar field will be strictly aligned with the vertical direction of 
illumination and will result in no modification of the resonance scattering polarization. The three inset images of Fig. 1 show the 
polarization degree at every point in the resolved disk and illustrate the loss of symmetry described at two different dipole field 
strengths. It can be seen that the polarization amplitude in the poles is no longer
compensated by the diminished polarization at the equator and the integral around the limb will not be zero. Explained in such 
simple terms, a dipole field results in measurable linear polarization from the star. 

To estimate the expected polarization degree from this effect we will use a spectral line arising from an  atomic system with angular
momenta $J_u=1$ and $J_l=0$ for the upper and lower levels respectively. This simplification allows us to use in our 
computations the analytic results of \cite{casini_hanle_2002} for this atomic system \footnote{A general transition or multi-level atom has been
described and can also be adequately computed \citep{landi_deglinnocenti_polarization_2004}.}. All lines that can be acceptably 
simulated by such an
atomic system are characterized by a critical Hanle field $B_c$, defined as the field at which the Zeeman splitting of the sublevels
 equals the natural width of the level  
\begin{equation}
 B_c = A/\omega_L = A/(8.79 g_U),
\end{equation}
 ($\omega_L$ being the Larmor frecuency of a 1G field, $A$ the 
Einstein coefficient of the level in MHz and $g_u$ the Land\'{e} factor of the upper level).
 With the dipole field strength put in terms of this critical Hanle field, the Hanle behavior
 of all such spectral lines is identical.
Fig. 1 shows the linear polarization of such lines in the presence of a dipole field integrated over the observable stellar disk. At zero
 field we are in the conditions of resonance scattering polarization that for symmetry reasons 
cancels out over the disk. As the magnetic field increases, depolarization at the equator happens but not at the poles, as described. 
Cancellation is not complete and a linear net polarization appears over the star. The effect grows for increasing field strengths and 
then saturates beyond the Hanle critical field. The reason is that Hanle effect can only take place whilst the Zeeman sublevels are not 
separated in energy more than their natural width. When the sublevels reach this critical field the quantum interference between the
 sublevels drops and the magnetic field can not further modify them. For reference, Fig. \ref{Bc} shows the value in G of the Hanle 
critical field for about 500 Hanle-sensitive lines in the visible and  near-IR spectral region (0.4 through 2.5 microns), extending previous 
compilations \cite{ignace_hanle_2001}. 
\begin{figure}[ht]
\resizebox{9cm}{!}{\includegraphics{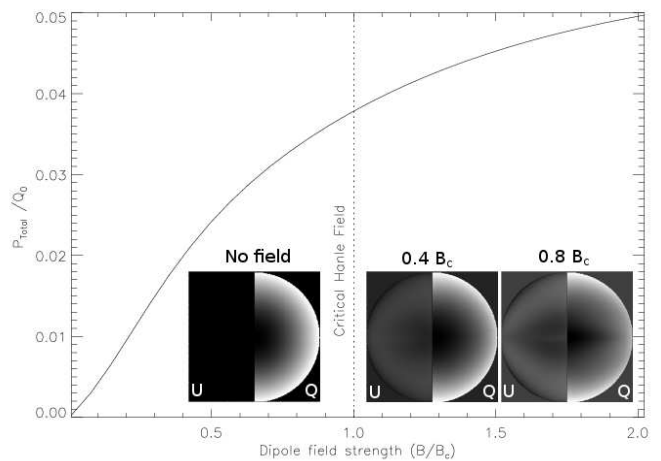}}
\caption{ Total polarization emitted from a star in a Hanle sensitive line  for a dipole field, aligned with the axis of rotation and
 included in the plane of the sky. Field strength of the dipole at the stellar surface is given in terms of the critical Hanle field of 
the observed line. The vertical axis is scaled in terms of the polarization $Q_0$ emitted by the  atom at 90-degree scattering in the 
absence of magnetic fields. The three inset images show the $Q$ and $U$ polarizations ($Q$ defined positive tangent to the local meridian)
in the disk-resolved star for 0, 50 and 100G of dipole field strength.}
\label{fig1}
\end{figure}

\begin{figure}[ht]
\resizebox{9cm}{!}{\includegraphics{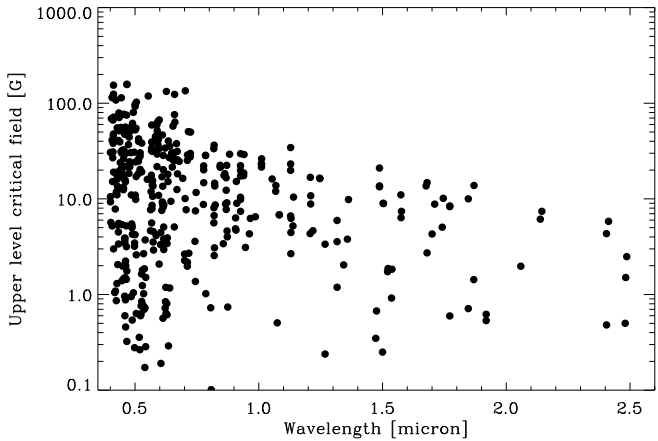}} 
\caption{ Critical Hanle field ($B_c$ in G) of about 500 lines sensitive to the Hanle effect in the visible and near-IR spectrum.}
\label{Bc}
\end{figure}

The computed polarization in Fig. 1 is given in terms of the zero-field polarization amplitude $Q_0$ of the line. Unfortunately the actual 
value of $Q_0$ is unknown for most (if not all) spectral lines. The reason for this is that it strongly depends on the actual anisotropy of
 the radiation field that illuminates the scattering atom. This anisotropy may depend on the formation height of the line in each 
stellar atmosphere, on the limb darkening conditions and on the pattern of intensities in the photosphere below the line. For example 
granulation in solar-like stars is known to influence this anisotropy and modifies the zero-field polarization $Q_0$ in these 
atmospheres \citep{trujillo_bueno_scattering_2007}. 

Although line-ratio techniques may sometimes overcome the problem of ignoring the value of $Q_0$, in general one is bound to estimate it by 
indirect reasonings. Examination of the solar spectra provides some clues on the value of $Q_0$. The scattering polarization of the visible spectra has 
been measured and is available in 
atlas \citep{gandorfer_second_2000}. Its spectral variation is very intriguing and so different from the intensity spectra that it is referred to 
as the \textit{second solar spectrum} \citep{stenflo_second_1997}. In it we can see that the \ion{Sr}{I} line at 460.7 nm, a paradigmatic 1-0 transition,   
presents a polarization level of 0.013 the intensity of the surrounding continuum. But this value is not the $Q_0$ amplitude we are 
searching. The solar photosphere is believed to be permeated by a turbulent and therefore random magnetic field that effectively 
depolarizes all Hanle-sensitive lines in the second-solar spectrum. The value of 0.013 corresponds to a Hanle-depolarized emission. 
The solar literature does not yet agree on how much depolarized the Hanle-sensitive lines are. Depending on the authors, mean field 
strengths from 20 to 150G are suggested, with the most precise studies \citep{trujillo_bueno_substantial_2004} pointing at the 
higher value. Depending on the assumptions, this may translate up to a depolarization of 80\% in the case 
of the Sr line. Therefore $Q_0$ is anywhere in the range of 0.013 to 0.06 times the intensity of the continuum.

Limb darkening is a major ingredient in the computation of the anisotropy of the radiation field, and in consequence of the actual value of 
$Q_0$. Since stars of different spectral classes present a variation in their limb darkening we have estimated the dependence of $Q_0$ with 
the spectral class. The result is a mere factor 3 of variation (from 0.005 through 0.015) between the coldest and hottest stars, assuming the line forms
in all those atmospheres, with higher $Q_0$ values for cooler stars. 

In conclusion, and albeit our ignorance of the actual value of the $Q_0$ zero-field polarization amplitude, we can safely 
assume it is in the range of a 
few percent. The proposed diagnostic results in integrated values (Fig.1) which are consequently of a few percent of a few percent, in 
other words at the level of $10^{-4}$ to $10^{-3}$ times the intensity of the line. This sets the feasability and the observing requirements 
for this diagnostic tool to be exploited.


\section{Oblique magnetic rotators and temporal modulation of the signal}
\begin{figure}[ht]
\resizebox{9cm}{!}{\includegraphics{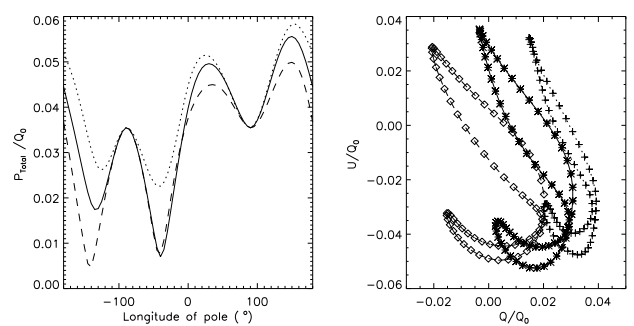}}
\caption{ Total linear polarization (in terms of $Q_0$)and plot of $Q$ vs. $U$ (defined with respect to a fixed direction in the sky corresponding to $U=0$ when 
perpendicular to the rotation axis) with varying rotational phase. The magnetic dipole strength has been fixed at  70\% the critical Hanle field, 
and the fixed astrographic latitude of the magnetic pole is of 30, 45 and 60 degrees for the dashed (crosses), continuous (stars) and dotted (diamond) 
lines respectively.}
\label{oblique}
\end{figure}

In general the magnetic dipole will not be aligned with the rotation axis of the star. The rotation of the dipole 
around the rotation axis will result in a temporal modulation of the observed linear polarization. This is very interesting since the shape of the
 temporal variation of polarization carries precious information on the orientation of the dipole in the star. To illustrate this 
we shall consider a star in which, as before, the rotation axis is on the plane of the sky and sets our preferred direction of 
polarization $Q$. A magnetic dipole of $B=0.7B_c$ is assumed with a constant astrographic latitude. It rotates around 
the rotation axis. Fig \ref{oblique} shows three cases of  modulation of the amplitude with rotational phase, and the angle of polarization 
(0 degrees meaning the preferred direction of polarization $Q$), for latitudes of 30, 45 and 60 degrees. In the description of the Hanle effect 
made above we mentioned that 
depolarization was accompanied by a rotation of the plane of polarization. Such rotation can be seen as a nonzero value of the Stokes $U$ 
polarization with non-zero magnetic field, and so is seen in the inset images of Fig. 1 where the Stokes U amplitude is plotted in the 
hemisphere of the disk at 0, 50 and 100G dipole field strength. The angle of polarization in Fig \ref{oblique} is defined as half the arctangent 
of the ratio of Q and U polarization amplitudes.

 Both the polarization amplitude and angle show a  strong assymetric behavior with rotational phase. The absence of symmetry with respect to the
central meridian of the star is due to the dependence of the polarization with twice the position angle, and not the angle itself.
Because of this lack of hemispheric symmetry, the signal from an oblique rotator should allow to determine with great precision 
(signal-to-noise levels allowing) the geometry of the magnetic dipole. In this aspect 
the proposed Hanle diagnostic outperforms similar diagnostics based upon the integrated circular polarisation, not just for its 
sensitivity to weaker fields, but also for the amount of topological information encoded in its temporal variation, assuming the dipole is 
not aligned with the rotation axis. It can be speculated that not just the relative angle between the rotation and dipolar axis can be 
ascertained, but also that the inclination of the rotation axis with respect to the plane of the sky can be determined, independently of the
rotation period, a result already pointed out by \cite{ignace_status_2001}.

\section{Spectral lines apt for the measurement and line addition techniques.}
Zeeman diagnostics of stellar fields made a huge leap forward when it was proposed that thousands of spectral lines accross the 
spectrum could be coherently added \citep{semel_zeeman-doppler_1996,donati_spectropolarimetric_1997,semel_multiline_2009}. Individual 
lines present circular polarisation levels below noise. But when 
many different lines are  appropriately added,  noise will add up incoherently diminishing in amplitude as the square root of the 
number of lines added with respect to the Zeeman signal. Signals meanwhile will add up coherently maintaining its amplitude. With this technique signal-to-noise ratios of $10^5$ 
are routinely achieved in many stars by the foremost instruments in this field: EsPadons and Narval \citep{donati_espadons:_2003,exhelle_2003,donati_espadons:new_2006,aurire_stellar_2003}. 

The amplitudes of linear polarization expected from the diagnostic proposed in this paper are of the level of $10^{-4}$ to $10^{-3}$ 
at its best. Its measurement requires long integration times, unless we can make use of a similar line-addition technique. The issue with the Hanle 
effect is that many lines do not form in regions where the anisotropy of the radiation field is sufficient to introduce resonance 
scattering polarization. Furthermore the actual level of zero-field polarization $Q_0$ is highly dependent on the quantum numbers of 
the transitions \citep{landi_deglinnocenti_polarization_2004}. Also it cannot be known a priori in what plane the polarization signal will be maximum, and
therefore both Q and U need to be measured at the telescope before being able to add the lines. Finally the last difficulty is that the spectral shape of the resonance scattering polarization is not
 standard for all lines, contrary to the Zeeman effect which induces a similar spectral shape of circular polarization to any 
Zeeman-sensitive line. All these are 
factors that make difficult adding lines in the Hanle case. \cite{belluzzi_spectroscopic_2009} examined and classified the spectral signatures and amplitudes of all lines in 
the second solar spectrum. The 1-0 atomic model we have used in the present work results systematically in a gaussian-like profile. 
\cite{belluzzi_spectroscopic_2009} refer to lines with that spectral signature as belonging to class S and identify a large total of 80 lines in the solar 
spectrum belonging to this class. From the point of view of stellar observations and line-addition techniques this means that in 
solar-like stars up to 80 lines, including the \ion{Sr}{I} line, are susceptible of being coherently added and a corresponding increase of a 
factor 9 in signal-to-noise ratio can be expected.  

Following the practice in the LSD treatment of Zeeman-Doppler Imaging \citep{brown_zeeman-doppler_1991} we can further propose that the lines of class S 
to be added can be weighted to improve their contribution to the integrated mean pseudo-line. Let us stress here an important point: In 
usual spatially-resolved Hanle diagnostics (e.g. solar applications) the lines present a linear polarization at zero magnetic field. 
Adding up lines in this scenario will just result in a dilution of the magnetic signal rather than a reinforcement. At first sight line
 addition would appear as counterproductive. But in the stellar non-resolved case, the integral for all position angles at zero field
 results in a cancellation of the zero-field signal. Therefore in the stellar observation, the Hanle effect is also characterized by 
a zero signal for zero field and non-zero signal for growing magnetic field. This unexpected fact is what permits proposing line 
addition: the individual line signals are all signatures of non-zero field. And in analogy to the weighting of lines by their Land\'e 
factor in the Zeeman case, it appears as judicious to weight lines by the quotient of the line polarizability over the critical 
Hanle field of each line.

In summary, the use of lines of the class S  opens the possibility to add lines coherently, and build up the signal-to-noise ratio in the measurement 
of linear polarization signature of disk-integrated Hanle effect of photospheric lines. We wish to leave this is a proposition for further work, since
many other aspects need to be worked out to put in practice any line addition scheme for Hanle signatures, aspects that are beyond the scope of the present work

\section{Interest as a diagnostic}

A dipolar magnetic field breaks the rotational symmetry in the direction of linear polarization from resonance scattering. An immediate
result is that a Hanle-induced linear polarization signal appears even in disk-integrated stellar observations. Its amplitude is 
directly related to the dipole field strength although highly influenced by the geometry of the field and its variation in time with respect 
to the observer. As with any Hanle diagnostics it is sensitive to very weak fields which opens new observational windows in stellar magnetism.
But, also as with any Hanle diagnostics, the expected signals are of low amplitude. Such low amplitudes set strong limitations 
in terms of signal-to-noise ratios for the observation. However, we estimate that they are not beyond the reach of present 
spectropolarimetric instruments.

The small amplitudes can be partially overcame by line-addition, similar to the Zeeman case. About 80 lines can be identified 
in the visible spectra of solar-type
stars ready for addition and potentially affording a factor 9 increase in the signal-to-noise ratio for fixed exposure times and instrument. 
Colder stars may present a higher number of lines ready for addition, including molecular lines. They also present an
enlarged anisotropy in the line formation due to its stronger limb darkening that results in stronger resonance scattering polarization.

The method works best for stars with simple dipolar topologies, that simplify the analysis. These fields are to be found in 
several stages of stellar evolution. Upper main sequence stars (with masses above 1.5M$_{\sun}$) are known to host them, and while
 Solar-type stars are dominated by fields with a very complicated geometry, mostly dipolar structures reappear in later stages of 
stellar evolution, in fully convective stars with spectral types later than M3 \citep{donati_magnetic_2009}.

Regarding field strengths, our method would prove useful in the study of several populations: Herbig Ae/Be stars, mostly in a 
pre-main sequence stage, have large scale fields in the range 0.5-5 kG \citep{alecian_magnetism_2009}. Into the main sequence, the few O stars with field 
measurements are $\theta^1$ Ori \citep{donati_magnetic_2002}, 
sporting a $\sim1$kG field, HD 19161, with $\sim200$G \citep{donati_discovery_2006} and the 
supergiant $\zeta$ Ori A, with a field within 50 to 100G \citep{bouret_weak_2008}; in fact, the Hanle effect will be one of the diagnostics 
used in the UV by FUSP (Fast Ultraviolet Spectro-Polarimeter, \cite{nordsieck_far-ultraviolet_1999}).
Peculiar Ap and Bp stars are known to have fields up to several tens of kG with a minimum around 300G 
\citep{aurire_discovery_2009}, and so they fall beyond our reach, but their
 normal A, B and F counterparts have no measurable fields to date, down to errors from 10 to 100G \citep{donati_magnetic_2009}. Our 
diagnosis could lower these detectability thresholds. Field intensity seems to decay with stellar type, and so G giants show fields 
with poloidal structure and strength of several tens to 100 G \citep{aurire_discovery_2009,lbre_lithium_2009}.

Although fields below 1G have been detected in Pollux, a 
K0II star \citep{aurire_discovery_2009}, little is known about more evolved giants. 
The detection of the 1G field in Pollux required the combination of more than 90 spectra taken during 18 months and with 
S/N$>1000$,  a prohibitive amount of observational time.
In contrast, having a tool sensitive enough to infer the properties of dipolar fields below the actual sensitivity limits 
 would provide very useful information on the evolution of magnetic 
fields. Particularly in high mass stars, as in these giants, the magnetic fluxes will be smaller than their dwarf counterparts. This
 population is of particular interest, as they are the progenitors of neutron stars, and thus necessary to get the whole picture of the 
evolution of magnetic fields.

\begin{acknowledgements}
A.A.R gratefully aknowledges financial support from the Consolider-GTC project and 
Spanish Ministerio de Ciencia e Innovaci\'{o}n (MICINN) under grant AYA2007-63881.
C.G.F. gratefully aknowledges financial support from the Consolider-GTC project and 
Spanish Ministerio de Ciencia e Innovaci\'{o}n (MICINN) under grant AYA2008-06166-C03-03.
\end{acknowledgements}

\bibliographystyle{/home/arturo/TeX/aanda/bibtex/aa}
\bibliography{/home/arturo/TeX/0biblio/aamnem99,/home/arturo/TeX/0biblio/biblio,/home/arturo/TeX/0biblio/articulos,/home/arturo/Dropbox/art41.bib}

\end{document}